\newcommand\apjcls{1}
\newcommand\aastexcls{2}
\newcommand\othercls{3}

\newcommand\papercls{\aastexcls}
\documentclass[tighten, times, twocolumn]{aastex7}  

\if\papercls \apjcls
\usepackage{apjfonts}
\else\if\papercls \othercls
\usepackage{epsfig}
\usepackage{margin}
\usepackage{times}
\fi\fi
\usepackage{ifthen}
\usepackage{natbib}
\usepackage{amssymb, amsmath}
\usepackage{appendix}
\usepackage{etoolbox}
\usepackage[T1]{fontenc}
\usepackage{paralist}
\everymath{\displaystyle}
\usepackage{url}

\if\papercls \apjcls
\newcommand\aas{\ref@jnl{AAS Meeting Abstracts}}
\newcommand\dps{\ref@jnl{AAS/DPS Meeting Abstracts}}
\newcommand\maps{\ref@jnl{MAPS}}
\else\if\papercls \othercls
\usepackage{astjnlabbrev-jh}
\fi\fi

\if\papercls \aastexcls
\hypersetup{citecolor=blue, 
            linkcolor=blue, 
            menucolor=blue, 
            urlcolor=blue}  
\else
\usepackage[
bookmarks=true,           
bookmarksnumbered=true,   
colorlinks=true,          
citecolor=blue,           
linkcolor=blue,           
menucolor=blue,           
urlcolor=blue,            
linkbordercolor={0 0 1},  
pdfborder={0 0 1},
frenchlinks=true]{hyperref}
\fi

\if\papercls \othercls

\else

\fi

\providecommand{\adsurl}[1]{\href{#1}{ADS}}

\makeatletter
\patchcmd{\NAT@citex}
  {\@citea\NAT@hyper@{%
     \NAT@nmfmt{\NAT@nm}%
     \hyper@natlinkbreak{\NAT@aysep\NAT@spacechar}{\@citeb\@extra@b@citeb}%
     \NAT@date}}
  {\@citea\NAT@nmfmt{\NAT@nm}%
   \NAT@aysep\NAT@spacechar\NAT@hyper@{\NAT@date}}{}{}

\patchcmd{\NAT@citex}
  {\@citea\NAT@hyper@{%
     \NAT@nmfmt{\NAT@nm}%
     \hyper@natlinkbreak{\NAT@spacechar\NAT@@open\if*#1*\else#1\NAT@spacechar\fi}%
       {\@citeb\@extra@b@citeb}%
     \NAT@date}}
  {\@citea\NAT@nmfmt{\NAT@nm}%
   \NAT@spacechar\NAT@@open\if*#1*\else#1\NAT@spacechar\fi\NAT@hyper@{\NAT@date}}
  {}{}
\makeatother

\makeatletter
\DeclareRobustCommand{\lowcase}[1]{\@lowcase#1\@nil}
\def\@lowcase#1\@nil{\if\relax#1\relax\else\MakeLowercase{#1}\fi}
\makeatother

\DeclareSymbolFont{UPM}{U}{eur}{m}{n}
\DeclareMathSymbol{\umu}{0}{UPM}{"16}
\let\oldumu=\umu
\renewcommand\umu{\ifmmode\oldumu\else\math{\oldumu}\fi}

\if\papercls \othercls

\else

\fi

\let\oldsim=\sim
\renewcommand\sim{\ifmmode\oldsim\else\math{\oldsim}\fi}
\let\oldpm=\pm
\renewcommand\pm{\ifmmode\oldpm\else\math{\oldpm}\fi}
\newcommand\by{\ifmmode\times\else\math{\times}\fi}

\newbox{\wdbox}
\renewcommand\c{\setbox\wdbox=\hbox{,}\hspace{\wd\wdbox}}
\renewcommand\i{\setbox\wdbox=\hbox{i}\hspace{\wd\wdbox}}
\newcommand\n{\hspace{0.5em}}

\newcount\timect
\newcount\hourct
\newcount\minct
\newcommand\now{\timect=\time \divide\timect by 60
         \hourct=\timect \multiply\hourct by 60
         \minct=\time \advance\minct by -\hourct
         \number\timect:\ifnum \minct < 10 0\fi\number\minct}

\catcode`@=11

\newcommand\comment[1]{}

\newcommand\commenton{\catcode`\%=14}

\renewcommand\math[1]{$#1$}
\newcommand\mathshifton{\catcode`\$=3}

\let\atab=&
\newcommand\atabon{\catcode`\&=4}

\let\oldmsp=\sp
\let\oldmsb=\sb
\def\sp#1{\ifmmode
           \oldmsp{#1}%
         \else\strut\raise.85ex\hbox{\scriptsize #1}\fi}
\def\sb#1{\ifmmode
           \oldmsb{#1}%
         \else\strut\raise-.54ex\hbox{\scriptsize #1}\fi}
\newbox\@sp
\newbox\@sb
\def\sbp#1#2{\ifmmode%
           \oldmsb{#1}\oldmsp{#2}%
         \else
           \setbox\@sb=\hbox{\sb{#1}}%
           \setbox\@sp=\hbox{\sp{#2}}%
           \rlap{\copy\@sb}\copy\@sp
           \ifdim \wd\@sb >\wd\@sp
             \hskip -\wd\@sp \hskip \wd\@sb
           \fi
        \fi}
\def\msp#1{\ifmmode
           \oldmsp{#1}
         \else \math{\oldmsp{#1}}\fi}
\def\msb#1{\ifmmode
           \oldmsb{#1}
         \else \math{\oldmsb{#1}}\fi}

\def\supon{\catcode`\^=7}

\def\subon{\catcode`\_=8}

\def\supsubon{\supon \subon}

\newcommand\actcharon{\catcode`\~=13}

\newcommand\paramon{\catcode`\#=6}

\comment{And now to turn us totally on and off...}

\newcommand\reservedcharson{ \commenton  \mathshifton  \atabon  \supsubon 
                             \actcharon  \paramon}

\catcode`@=12
\reservedcharson

\if\papercls \apjcls

\else

\fi

\newcommand\chisq{\ifmmode{\chi\sp{2}}\else\math{\chi\sp{2}}\fi}
\newcommand\redchisq{\ifmmode{ \chi\sp{2}\sb{\rm red}}
                    \else\math{\chi\sp{2}\sb{\rm red}}\fi}
\newcommand\Teq{\ifmmode{T\sb{\rm eq}}\else$T$\sb{eq}\fi}
\newcommand\mjup{\ifmmode{M\sb{\rm Jup}}\else$M$\sb{Jup}\fi}
\newcommand\rjup{\ifmmode{R\sb{\rm Jup}}\else$R$\sb{Jup}\fi}
\newcommand\msun{\ifmmode{M\sb{\odot}}\else$M\sb{\odot}$\fi}
\newcommand\rsun{\ifmmode{R\sb{\odot}}\else$R\sb{\odot}$\fi}
\newcommand\mearth{\ifmmode{M\sb{\oplus}}\else$M\sb{\oplus}$\fi}
\newcommand\rearth{\ifmmode{R\sb{\oplus}}\else$R\sb{\oplus}$\fi}

\shorttitle{Disk truncation triggers relativistic jet launching in 3C120}
\shortauthors{Noda {\em et al.}}

\begin{document}

\title{Disk truncation triggers relativistic jet launching of a highly accreting supermassive black hole}

\author{Hirofumi~Noda}
\affiliation{Astronomical Institute, Tohoku University, 6-3 Aramakiazaaoba, Aoba-ku, Sendai, Miyagi 980-8578, Japan}
\email[show]{hirofumi.noda@astr.tohoku.ac.jp}  

\author{Kazuhiro~Hada}
\affiliation{Graduate School of Science, Nagoya City University, Yamanohata 1, Mizuho-cho, Mizuho-ku, Nagoya, Aichi 467-8501, Japan}
\affiliation{Mizusawa VLBI Observatory, National Astronomical Observatory of Japan, 2-12 Hoshigaoka, Mizusawa, Oshu, Iwate 023-0861, 
Japan}
\email[show]{hada@nsc.nagoya-cu.ac.jp}  

\author{Hiroto~Maruyama}
\affiliation{Graduate School of Science, Nagoya City University, Yamanohata 1, Mizuho-cho, Mizuho-ku, Nagoya, Aichi 467-8501, Japan}
\email[]{}  

\author{Taishu~Kayanoki}
\affiliation{Faculty of Engineering, University of Miyazaki, 1-1 Gakuen Kibanadai Nishi, Miyazaki, Miyazaki 889-2192, Japan}
\email[]{}  

\author{Kenji~Toma}
\affiliation{Frontier Research Institute for Interdisciplinary Sciences, Tohoku University, 6-3 Aramakiazaaoba, Aoba-ku, Sendai, Miyagi 980-857, Japan}
\affiliation{Astronomical Institute, Tohoku University, 6-3 Aramakiazaaoba, Aoba-ku, Sendai, Miyagi 980-8578, Japan}
\email[]{}  

\author{Taiki~Kawamuro}
\affiliation{Department of Earth and Space Science, The University of Osaka, 1-1 Machikaneyama, Toyonaka, Osaka 560-0043, Japan}
\email[]{}  

\author{Yoshiyuki~Inoue}
\affiliation{College of Systems Engineering and Science, Shibaura Institute of Technology, 307 Fukasaku, Minuma-ku, Saitama City, Saitama 337-8570, Japan}
\email[]{}  

\author{Tomohisa~Kawashima}
\affiliation{National Institute of Technology, Ichinoseki College, Takanashi, Hagisho, Ichinoseki, Iwate 021-8511, Japan}
\email[]{}  

\author{Soyeon~Yeo}
\affiliation{Department of Astronomy, Yonsei University, Yonsei-ro 50, Seodaemun-gu, Seoul 03722, Republic of Korea}
\email[]{}  

\author{Ilje~Cho}
\affiliation{Department of Astronomy and Atmospheric Sciences, Kyungpook National University, 80 Daehak-ro, Buk-gu, Daegu 41566, Republic of Korea}
\affiliation{Korea Astronomy and Space Science Institute, Daedeok-daero 77, Yuseong-gu, Daejeon 34055, Republic of Korea}
\email[]{}  

\author{Keiichi~Asada}
\affiliation{Institute of Astronomy and Astrophysics, Academia Sinica, 11F of Astronomy, Mathematics Building, AS/NTU No. 1, Sec. 4, Roosevelt Road, Taipei 10617, Taiwan}
\email[]{}  

\author{Yasushi~Fukazawa}
\affiliation{Department of Physics, Graduate School of Advanced Science and Engineering, Hiroshima University, 1-3-1 Kagamiyama, Higashi-Hiroshima, Hiroshima 739-8526, Japan}
\email[]{}  

\author{Yoshihiro~Ueda}
\affiliation{Department of Astronomy, Kyoto University, Kitashirakawa-Oiwake-cho, Sakyo-ku, Kyoto 606-8502, Japan}
\email[]{}



\begin{abstract}
Relativistic jets are among the most energetic phenomena in the Universe and influence galaxy evolution through energetic feedback, yet the physical connection between accretion flows and jet launching remains unresolved.
Jet production is widely attributed to the Blandford--Znajek mechanism, which extracts black hole rotational energy via large-scale magnetic fields sustained by an accretion flow. 
Although this mechanism is observationally well supported for hot, geometrically-thick accretion flows, highly accreting environments with strong jets present a paradox, as they are typically dominated by cold, geometrically-thin disks incapable of sustaining large-scale magnetic fields.
Here, we report coordinated X-ray microcalorimeter spectroscopy achieved by X-ray Imaging and Spectroscopy Mission (XRISM) and millimeter/radio interferometric observations by Global Millimeter VLBI Array (GMVA), Very Long Baseline Array (VLBA), and East Asian VLBI Network (EAVN) on the broad-line radio galaxy 3C120.
High-resolution X-ray spectroscopy reveals a relativistically-broadened Fe-K$\alpha$ line, indicating that the cold, geometrically-thin disk is truncated at $\sim 20~R_{\rm g}$ ($R_{\rm g}$ is the gravitational radius), with the inner region replaced by a hot, geometrically-thick flow. 
Concurrently, radio imaging indicates that jet profile extrapolation toward the black hole horizon yields a radius comparable to or even narrower than the disk truncation radius.
These results demonstrate that, even at high accretion rates, jet launching is linked to a hot inner flow, possibly via the Blandford--Znajek process, providing evidence for a universal disk--jet connection via geometric transition. 
\end{abstract}

\section{Introduction}

Active galactic nuclei (AGN), powered by accretion onto supermassive black holes, often launch relativistic jets: energetic, highly collimated plasma outflows traveling at near-light speeds. 
These jets are a major channel of cosmic feedback, influencing the evolution of their host galaxies and the surrounding intracluster medium. 
Recent millimeter-wavelength Very Long Baseline Interferometry (VLBI), most notably the Event Horizon Telescope (EHT) observations of the nearby, weakly accreting AGN M87, has provided unprecedented constraints on the innermost jet structure and its launch region~\citep{EHT2021}. 
Combined with general relativistic magnetohydrodynamic (GRMHD) simulations~(e.g., \citealt{Tchekhovskoy2011}), these pioneering observations have offered a prevailing paradigm in which powerful jets are produced by hot, geometrically-thick accretion flows that can accumulate and sustain strong magnetic flux near the black hole, enabling the Blandford--Znajek mechanism~\citep{BZ1977}. 
At high accretion rates, however, a critical gap remains. 
Standard accretion-disk theory predicts relatively cold, geometrically-thin disks in this regime, while numerical models suggest that such disks have difficulty transporting and maintaining the strong magnetic flux required for powerful jet production in the immediate vicinity of the black hole~\citep{Shakura1973,Lubow1994}. 
Reconciling powerful relativistic jets with such high-accretion-rate environments therefore remains a challenge in high-energy astrophysics. 

The broad-line radio galaxy 3C120 ($z=0.033$), with a black hole mass of $M_{\rm BH} \sim 6.3 \times 10^7M_{\odot}$\citep{Hlabathe2020}, provides an ideal laboratory for addressing this problem. 
Unlike weakly accreting AGNs such as M87 ($L/L_{\rm Edd} \sim 10^{-6}$, where $L_{\rm Edd}$ is the Eddington luminosity), 3C120 accretes at a relatively high rate, with $L/L_{\rm Edd} \sim 0.1$--$0.2$~\citep{Zdziarski2022}, while simultaneously launching a powerful relativistic jet. 
Long-term X-ray and radio monitoring has revealed that X-ray dips, interpreted as changes in the inner accretion flow, are followed by the emergence of bright superluminal knots in the radio jet \citep{Marscher2002}. 
X-ray spectral modeling of such events has suggested that these dips may correspond to a temporary truncation or instability of the innermost accretion disk, transferring energy and matter into the jet base~(e.g., \citealt{Lohfink2013}).
However, whether the inner disk edge truly undergoes structural reconfiguration has remained controversial. 
This is largely because previous X-ray observations lacked the spectral resolution to unambiguously decouple the intrinsic relativistic Fe-K$\alpha$ line profile from both the continuum and complex absorption/emission features produced by disk winds~\citep{Tombesi2014,Tombesi2017}, as well as narrow components from distant reflection. 
Furthermore, previous VLBI observations were limited to imaging regions relatively far downstream from the jet-launching site, leaving the precise geometry of the disk--jet interface unresolved.

To investigate this disk--jet connection, we performed coordinated multiwavelength X-ray and VLBI observations of 3C120. 
We observed 3C120 with the X-Ray Imaging and Spectroscopy Mission (XRISM; \citealt{Tashiro2025}) from 19 to 24 March 2025 (UTC) as part of the Cycle 1 Guest Observation program. 
Contemporaneously, we obtained millimeter- and radio-VLBI observations with the Global Millimeter VLBI Array (GMVA) at 86~GHz, the Very Long Baseline Array (VLBA) at 43 and 15~GHz, and the East Asian VLBI Network (EAVN) at 22~GHz. 
The high-resolution X-ray spectroscopy provided by XRISM enables us to constrain the geometry of the innermost accretion flow through the relativistic Fe-K$\alpha$ emission, while the VLBI observations trace the structure and collimation of the innermost jet. 
By combining these complementary probes of the accretion flow and jet, we obtain geometric constraints on the disk--jet interface in a highly accreting AGN and provide new observational insight into how powerful relativistic jets are launched in this regime.

This paper is organized as follows. 
In Section~2, we describe the XRISM and VLBI observations of 3C120, together with the data reduction procedures for both the X-ray and VLBI data. 
In Section~3, we present the spectral and imaging analyses based on the XRISM and VLBI data, respectively.  
In Section~4, we discuss the implications of our results for a universal jet-launching engine in highly accreting black holes. 
A comprehensive summary of our observational findings and the resulting physical picture of the disk--jet interface is schematically illustrated in Fig.~\ref{main}.
We adopt the cosmological parameters $H_0 = 73$~km~s$^{-1}$~Mpc$^{-1}$, $\Omega_{\Lambda} = 0.73$, and $\Omega_{\rm m} = 0.27$ throughout this paper.

\section{Observation and Data Reduction}
\label{two}

\subsection{XRISM Observations and Data Reduction}

XRISM observed 3C120 from 19 to 24 March 2025 (UTC) during the Cycle 1 Guest Observer program (OBSID 201019010), providing a net exposure of $\sim200$~ks. 
Data reduction was performed using the HEASOFT software package (v6.36) and the CALDB released in November 2025.

For the Resolve microcalorimeter \citep{Kelley2025}, we selected only high-resolution primary (Hp) grade events for analysis. 
The source spectrum was extracted by integrating data from all pixels except for Pixel 27, which was excluded to mitigate large gain fluctuations. 
Redistribution Matrix Files (RMFs) were generated using the \texttt{rslmkrmf} task with the \texttt{whichrmf=X} option. 
Ancillary Response Files (ARFs) were produced using \texttt{xaarfgen} based on exposure maps created with \texttt{xaexpmap}. 
The Non-X-ray Background (NXB) was estimated via \texttt{rslnxbgen}, and the resulting model was included as a background component in the spectral fitting. 
Because the gate valve was closed during this observation, the energy range was restricted to 3.0--12.0~keV. 
The Resolve spectrum was binned using the optimal binning algorithm via \texttt{ftgrppha} and fitted in XSPEC using $C$-statistics.

For the Xtend CCD \citep{Noda2025}, data were acquired in 1/8-window mode and reduced following standard pipelines. 
The source spectrum was extracted from a $5'\times2.3'$ rectangular region centered on the source, with a background spectrum obtained from an equivalent off-source region. 
To maintain consistency with the Resolve analysis and to minimize systematic uncertainties associated with the calibration in 1/8-window mode at lower energies, we restricted the energy range to 3.0--12.0~keV. 
This choice also ensures that our spectral modeling is not biased by additional complex components at lower energies, such as the soft X-ray excess and/or jet-related emission \citep{Kataoka2007, Lohfink2013}, which are beyond the scope of this study focusing on the Fe-K band.
RMF and ARF files were generated using \texttt{xtdrmf} and \texttt{xaarfgen}, respectively. 
The Xtend spectrum was binned to ensure a minimum of 20 counts per bin and analyzed in XSPEC using $C$-statistics. 

For visualization in all figures, Resolve and Xtend spectra were further rebinned using the setplot rebin command.
For the X-ray analysis, unless otherwise stated, the tables and figures use $1\sigma$ error ranges. 

\subsection{VLBI Millimeter and Radio Observations}

On 26--27 April 2025, we made GMVA observations of 3C120 at 86\,GHz. 
A total of 18 stations participated in the session worldwide: five stations in Europe (Effelsberg, Onsala, Yebes, Mets\"{a}hovi, Greenland Telescope), eight stations of VLBA (Brewster, Fort Davis, Kitt Peak, Los Alamos, Mauna Kea, North Liberty, Owens Valley, Pie Town), 4 stations of the Korean VLBI Network (KVN; Yonsei, Ulsan, Tamna, Pyeongchang) and one station in Chile (APEX). 
The configuration provided maximum baseline lengths exceeding 9500\,km in both the east-west and north-south directions. 
The observations were carried out in full-track mode over a continuous 20-hour period from 9 UT on April 26 to 5 UT on April 27. Data were recorded at a rate of 4\,Gbps, using 2-bit sampling and a total bandwidth of 512\,MHz per polarization, divided into eight 64\,MHz subbands. 
Short scans of the bright compact calibrator J0423--0210 was interleaved with the primary scans of 3C120. 
The raw data were correlated with the DiFX correlator at the Max Planck Institute for Radio Astronomy, and subsequent data calibration was performed using the National Radio Astronomy Observatory (NRAO) Astronomical Image Processing System (AIPS~\citealt{Greisen2003}) following standard GMVA data reduction procedures~\citep{Marti-vidal2012}. 
Part of VLBA stations (e.g., Los Alamos and Pie Town) yielded weak or no fringe detections due to severe weather conditions or insufficient antenna pointing. 
Nevertheless, the overall data quality throughout the session was sufficient for imaging thanks to the robust fringe detections across the remaining stations, providing dense baseline coverage from 10\,M$\lambda$ to 2600\,M$\lambda$. 
After the initial calibration, the data were averaged over all eight subbands for imaging. Imaging was performed in DIFMAP~\citep{Shepherd1997} using iterative CLEAN and phase/amplitude self-calibration. We adopted a uniform weighting scheme to maximize the contribution of long baseline data and achieve high angular resolution. The resulting synthesized beam was quasi-circular with a half-power beam width (HPBW) of 42--60 $\mu$as, and the final image rms noise level reached $1$\,mJy\,beam$^{-1}$ near the source (i.e., dynamic-range limited). 

In addition to the GMVA data, we also analyzed contemporaneous lower-frequency VLBI data of 3C120 obtained with VLBA and EAVN to extend measurements of jet collimation profile to larger distances downstream. For VLBA, we utilized publicly available calibrated datasets from the Boston University BEAM-ME program  (43\,GHz; \citealt{Jorstad2017}) and the Monitoring of Jets in Active galactic nuclei with VLBA Experiments (MOJAVE) program (15\,GHz; \citealt{Lister2018}). 
For both BEAM-ME and MOJAVE, we produced stacked images by combining multiple closely spaced epochs (2025 Feb 16, Mar 9, Mar 23 and Apr 14 for BEAM-ME; and Feb 21, Mar 16 and Jun 1 for MOJAVE) in order to enhance sensitivity to the extended jet emission at each frequency. 
For EAVN, we conducted observations at 22\,GHz on 23 April 2025 as part of our ongoing monitoring program of this source (Maruyama et al. in prep.), where a total of 10 stations joined across East Asia (Ulsan, Tamna, Pyeongchang, Mizusawa, Iriki, Ogasawara, Ishigaki, Takahagi and Nanshan). The EAVN data were correlated at the Daejeon GPU correlator installed at Korea Astronomy and Space Science Institute. Initial calibration of the correlated EAVN data was performed in AIPS following standard EAVN data reduction procedures. For all lower-frequency VLBI datasets, final imaging was performed in DIFMAP using the CLEAN algorithm. 
All uncertainties quoted in the radio VLBI analysis represent the $1\sigma$ error range.

\section{Analysis}
\subsection{X-ray Spectral Analysis}

\begin{figure*}[t]
\begin{center}
\includegraphics[width=13cm, angle=0]{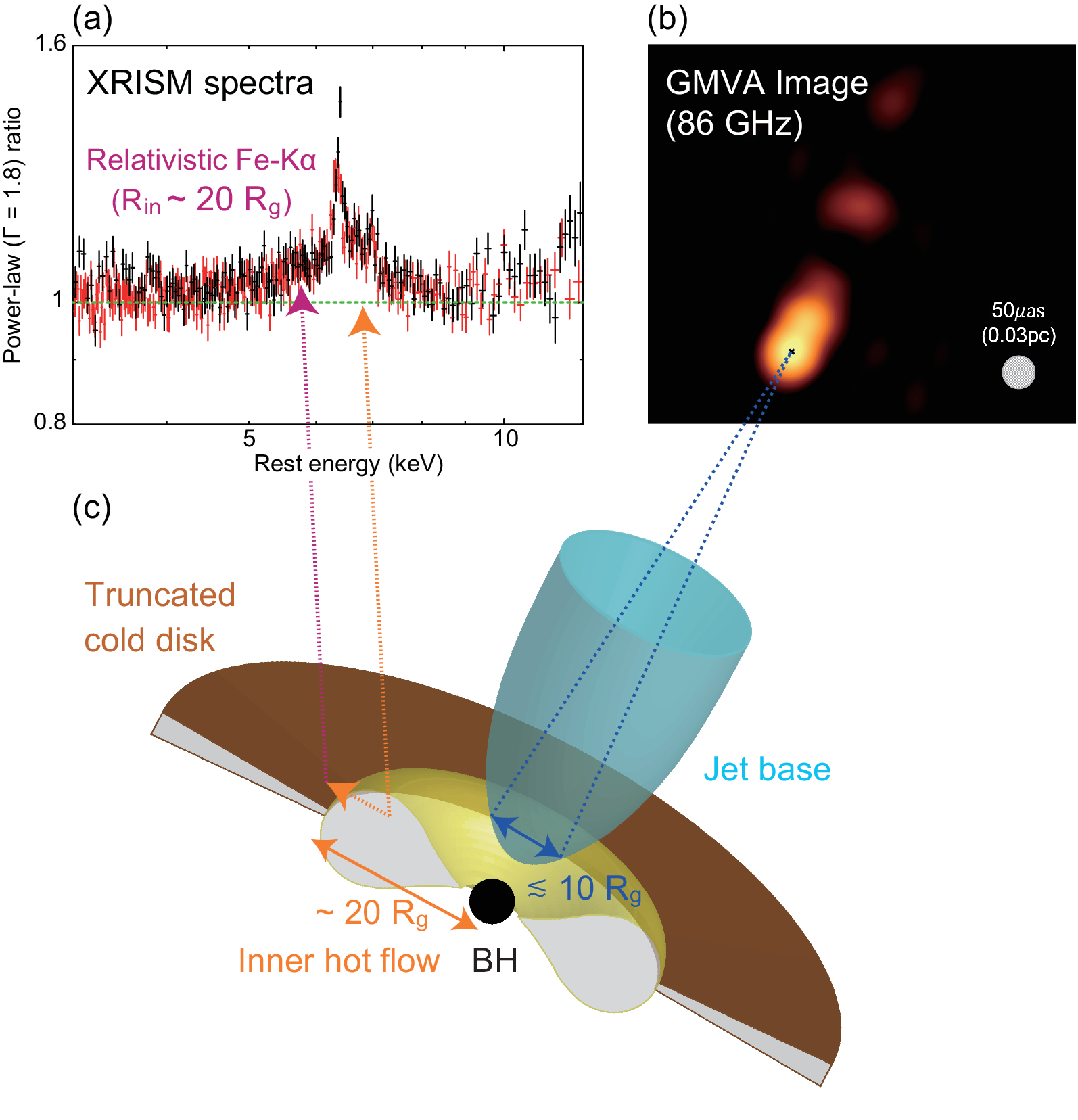}
\end{center}
\vspace{-0.5cm}
\caption{(a) X-ray spectra of 3C120 obtained by the XRISM Resolve microcalorimeter (black) and Xtend CCD (red). Data are shown as ratios to a power-law continuum with a photon index of $\Gamma = 1.8$. 
(b) High-resolution GMVA 86~GHz image of the relativistic jet base in 3C120. 
(c) Schematic illustration of the inferred accretion flow and jet structure near the black hole in 3C120. The cold, geometrically-thin disk (brown) is truncated at approximately $20~R_{\rm g}$. Within this radius, the flow transitions to a hot, geometrically-thick accretion flow (yellow), which inwardly transport large-scale strong magnetic fields to power and collimate the relativistic jet (cyan) efficiently. The jet is launched from within $10~R_{\rm g}$ and is subsequently collimated into a parabolic profile.
}
\label{main}
\end{figure*}

\begin{figure*}[t]
\begin{center}
\includegraphics[width=15cm, angle=0]{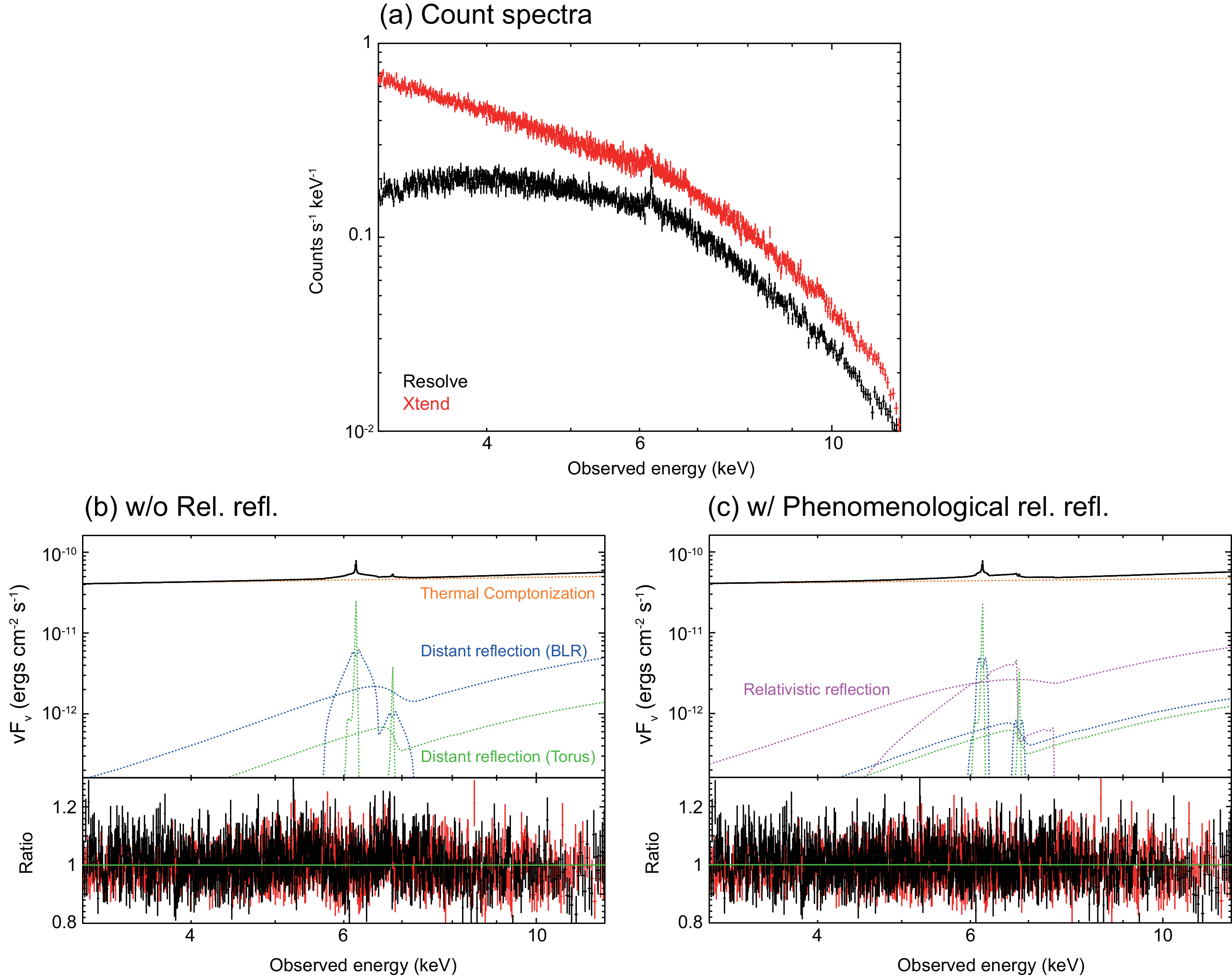}
\end{center}
\caption{(a) XRISM Resolve (black) and Xtend (red) spectra of 3C120 in count space. (b) The top panel shows the best-fit model using a baseline model excluding any relativistic reflection component, and the bottom panel shows the data-to-model ratios. (c) Same as panel (b), but including a phenomenological relativistically-blurred reflection component.
In panels (b) and (c), the total model (black solid line) comprises the primary thermal Comptonization continuum (orange), the relativistically-blurred reflection (magenta), and neutral Fe-K$\alpha$ emission lines (intermediate-width, blue; narrow, green) originating from distant reflectors. The Galactic photoelectric absorption effect is incorporated in the spectral modeling. }
\label{XRISMfit}
\end{figure*}

\renewcommand{\arraystretch}{1.0}
\begin{table*}
 \caption{Parameters of the simultaneous fits to the Resolve and Xtend spectra of 3C120 by the model without and with the physical relativistic reflection component. All quoted uncertainties represent the $1\sigma$ confidence level. Uncertainties for the model without relativistic reflection are omitted because the fit is unacceptable.}
 \vspace{0.5cm}
 \begin{center}
 \begin{tabular}{cccc}

  \hline\hline
   Model & Parameter & w/o Rel. refl. & w/ Rel. refl. \\\hline
   \texttt{TBabs} & $N_{\rm H,Gal}^{\rm a}$ & \multicolumn{2}{c}{1.02 (fix)} \\[1.5ex]
   
    \texttt{zpowerlw} & $\Gamma$ & $1.85$ &  $1.83 \pm 0.01$    \\
    (Thermal Compton) & $N_{\rm PL}^{\rm b}$ & $2.3$ & $ 2.0 \pm 0.1$ \\[1.5ex]	
	
    \texttt{relxill} & $q_{\rm Rel}$ & $-$ & $4.6^{+5.4L}_{-1.3}$  \\
   (Relativistic reflection)  	& $i_{\rm Rel}$ (degrees) & $-$ & $31.3^{+1.0}_{-0.4}$   \\
    			& $R_{\rm in, Rel}~(R_{\rm g})$ & $-$ & $17.4^{+6.3}_{-6.5}$  \\
    			& $A_{\rm Fe}$~(Solar) & $-$ & $1.0^{+1.0}_{-0.2}$  \\
			& $\log \xi$ & $-$ & $2.96^{+0.09}_{-0.06}$ \\
    			 & $N_{\rm Rel}^{\rm c}$ & $-$ & $ 8.73^{+1.75}_{-1.45}$ \\[1.5ex]
			 
    \texttt{rdblur*MYTorus$_{\rm BLR}$} & $q_{\rm BLR}$ & $2.5$ & $2.7^{+2.0}_{-1.2}$ \\
    			(Broad-line region) & $R_{\rm in, BLR}~(R_{\rm g})$ & $100^P$  & $824^{+295}_{-149}$ \\
			& $i_{\rm BLR}$ (degrees) & $ 45^P $  &  $= i_{\rm Rel}$   \\
    			 & $N_{\rm BLR}^{\rm d}$ & $16.6$ & $5.4^{+1.2}_{-0.8}$ \\
			 & EW$_{\rm BLR}$~(eV) & $67$ & $24^{+3}_{-8}$ \\[1.5ex] 	
						
    \texttt{rdblur*MYTorus$_{\rm Torus}$} & $q_{\rm Torus}$ & \multicolumn{2}{c}{$2$ (fix)}  \\
    			(Dusty torus) & $R_{\rm in, Torus}~(10^4~R_{\rm g})$ & $ 1^P $  & $1.12^{+25.38}_{-0.12L}$ \\
			& $i_{\rm Torus}$ (degrees) & $= i_{\rm BLR}$  & $= i_{\rm Rel}$ \ \\
    			 & $N_{\rm Torus}^{\rm d}$ & $4.7$ & $3.3^{+0.5}_{-0.9}$ \\
			 & EW$_{\rm Torus}$~(eV) & $18$ & $14^{+5}_{-4}$ \\\hline

	\multicolumn{2}{c}{$C$-statistic/d.o.f.}  & 4209.0/4157 & 4106.2/4152\\
	\multicolumn{2}{c}{AIC} & 4225.0 & 4132.3 \\ 		     
  \hline\hline
  \end{tabular}
\smallskip
\begin{flushleft}
\small
\textbf{Note:} 
Superscript and subscript $L$ indicate that the parameter confidence interval reached the hard limit. Superscript $P$ indicates that the parameter was pegged at the assumed boundary.\\
$^{\rm a}$ Equivalent hydrogen column density in units of $10^{21}$~cm$^{-2}$.\\
$^{\rm b}$ Normalization of \texttt{zpowerlw} in units of $10^{-2}$~photons~keV$^{-1}$~cm$^{-2}$~s$^{-1}$ at 1~keV.\\
$^{\rm c}$ Normalization of \texttt{relxill} in units of $10^{-5}$.\\
$^{\rm d}$ Normalization of \texttt{MYTorus} in units of $10^{-3}$.\\
\end{flushleft}
\end{center}
\label{tab1}
\end{table*}

\begin{figure*}[t]
\begin{center}
\includegraphics[width=15cm, angle=0]{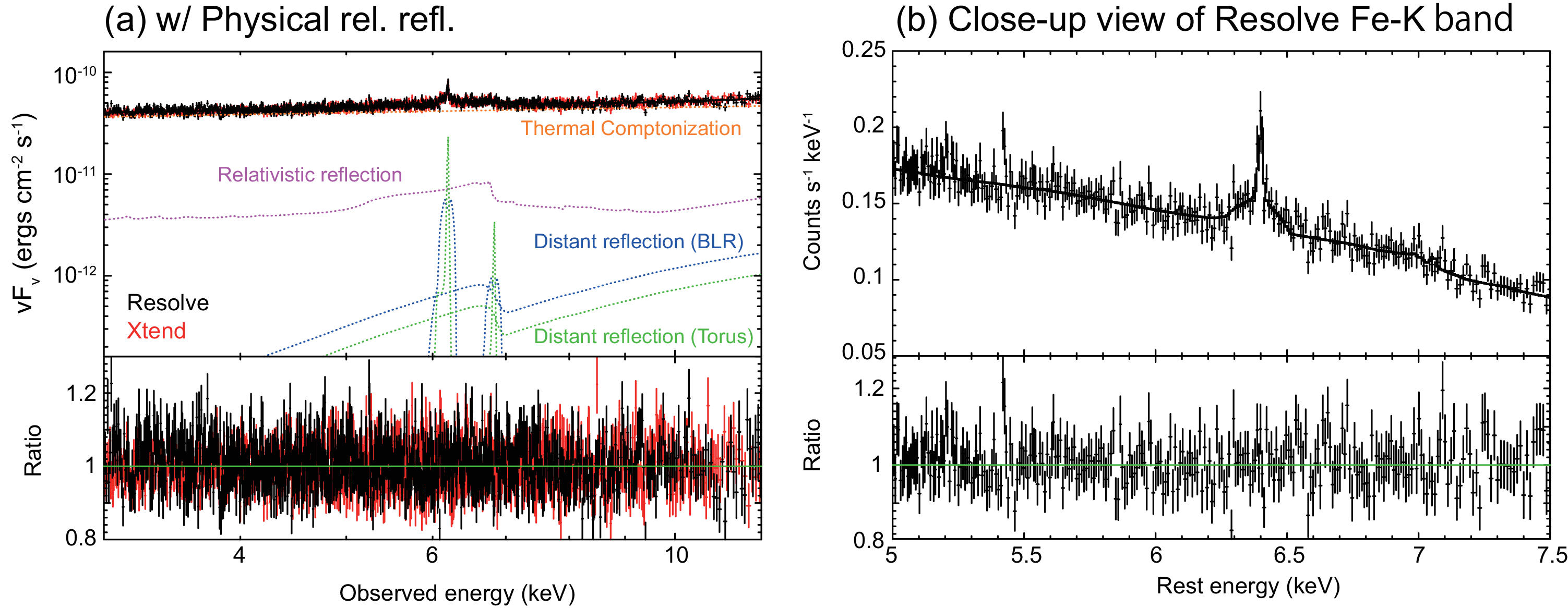}
\end{center}
\caption{(a) XRISM Resolve (black) and Xtend (red) spectra of 3C120 in the observed frame, shown together with the best-fit relativistic reflection model. The total model consists of a primary thermal Comptonization continuum (orange), a relativistically-blurred reflection component (magenta), and narrow neutral Fe-K$\alpha$ emission lines from distant reflectors, namely the broad-line region (BLR; blue) and dusty torus (green). (b) Close-up view of the XRISM Resolve spectrum and the best-fit total model in the Fe-K region, corrected to the rest frame and shown in units of counts~s$^{-1}$~keV$^{-1}$ instead of $\nu F_{\nu}$. In both panels, the bottom subpanel shows the data-to-model ratios.}
\label{Xray}
\end{figure*}

Figure~\ref{main}(a) shows the ratios of the Resolve X-ray microcalorimeter and Xtend CCD spectra to a power-law continuum with a photon index of $\Gamma=1.8$, which represents the primary thermal Comptonization component. 
This model-independent approach clearly reveals broad Fe-K$\alpha$ emission features extending from approximately 4~keV to 7~keV, in addition to a narrow neutral Fe-K$\alpha$ line at 6.4~keV. 
To examine this structure, we simultaneously fitted the Resolve and Xtend spectra in the 3.0--12.0~keV energy band (Fig.~\ref{XRISMfit}a), using the steps described below.

Initially, we employed a baseline model without a relativistically blurred reflection component, defined as $\texttt{TBabs}*( \texttt{zpowerlw} + \sum_{j=1,2} \texttt{rdblur}_{j}*(\texttt{MYtorusL}_j + \texttt{MYtorusS}_j) )$. 
The \texttt{TBabs} component accounts for Galactic photoelectric absorption, with the Galactic column density fixed at $N_{\rm H, Gal} = 1.02 \times 10^{21}$~cm$^{-2}$ \citep{HI4PI2016}. 
Intrinsic neutral absorption was not included in our model, as its effect is negligible within this specific energy band. 
The primary continuum was modeled using \texttt{zpowerlw}, where the photon index ($\Gamma$) and normalization ($N_{\rm PL}$) were treated as free parameters, and the redshift was fixed at $z = 0.033$.

To account for the narrow Fe-K$\alpha$ lines and their associated Compton humps, we included two sets of $\texttt{rdblur}*\left(\texttt{MYtorusL} + \texttt{MYtorusS}\right)$ components \citep{Fabian1989, Murphy2009, Yaqoob2024}, representing the broad-line region (component BLR) and the dusty torus (component Torus). 
The \texttt{rdblur} kernel was employed to approximate the line broadening in each environment. 
For both components, the column densities were fixed at $N_{\rm H, MY} = 1 \times 10^{24}$~cm$^{-2}$, as they could not be significantly constrained by the data. 
The inclination angle ($i_{\rm MY}$) was tied between the BLR and Torus components and allowed to vary. 
The inner radii ($R_{\rm in, BLR}$, $R_{\rm in, Torus}$) and normalizations ($N_{\rm BLR}$, $N_{\rm Torus}$) were treated as independent free parameters between the two sets, while remaining tied within each individual set. 
The emissivity index ($q$), where the emissivity profile is assumed to scale $\propto r^{-q}$, was allowed to vary
for the BLR component ($q_{\rm BLR}$), whereas that for the Torus component was fixed at $q_{\rm Torus} = 2$ to represent a distant reprocessor. 
To distinguish this emission from strongly relativistic disk reflection, we imposed a lower limit of $100$~$R_{\rm g}$ on $R_{\rm in, BLR}$, where $R_{\rm g} = \frac{GM_{\rm BH}}{c^2}$ is the gravitational radius. 
Similarly, a lower limit of $10^4$~$R_{\rm g}$ was applied to $R_{\rm in, Torus}$ to avoid unphysically broad profiles. 
Furthermore, the inclination was restricted to $i_{\rm MY} \le 45$ degrees, consistent with the Type-1 radio galaxy classification of 3C120 and its observed one-sided jet. 
The outer radii ($R_{\rm out, MY}$) were fixed at $10^4~R_{\rm g}$ and $10^7~R_{\rm g}$ for the BLR and Torus components, respectively. 
The photon indices for both components were tied to the $\Gamma$ of the primary continuum.
This initial fit was statistically unacceptable with $C$-statistic/d.o.f. $= 4209.0/4157$ and Akaike Information Criteria (AIC; \citealt{Akaike1974}) $=4225.0$, leaving prominent residuals in the $5$--$7$~keV range (Fig.~\ref{XRISMfit}b). 
Notably, $i_{\rm BLR}$ and $R_{\rm in, BLR}$ (component BLR) pegged at their respective boundaries of 45 degrees and $100$~$R_{\rm g}$ in an attempt to compensate for the unmodeled broad emission feature.

To phenomenologically test for relativistic broadening, we added a third $\texttt{rdblur} * (\texttt{MYtorusL} + \texttt{MYtorusS})$ set (component Disk), allowing the inner radius $R_{\rm in, Disk}$ to vary below $100$~$R_{\rm g}$. 
The fitted model at this stage was $\texttt{TBabs}*(\texttt{zpowerlw} + \sum_{j=1,2,3} \texttt{rdblur}_j *(\texttt{MYtorusL}_j + \texttt{MYtorusS}_j))$. 
The inclination angle $i_{\rm MY}$ was tied across all reflection components, and the outer radius $R_{\rm out, Disk}$ for component Disk was fixed at $1000$~$R_{\rm g}$. 
This addition significantly improved the fit, yielding $C$-statistic/d.o.f. $=4124.6/4154$ and AIC $= 4146.7$. 
As shown in Fig~\ref{XRISMfit}(c), the residuals observed in the baseline model were successfully eliminated. 
A $\Delta\text{AIC}$ of $78.3$ further confirmed the statistical necessity of a relativistically blurred component. 
The resulting inner disk radius was $R_{\rm in, Disk} = 10.9^{+4.1}_{-1.9}$~$R_{\rm g}$, while the inclination angle remained relatively large at $i_{\rm MY} = 41.6 \pm 0.3$ degrees (both at the 1$\sigma$ confidence level).

Finally, we replaced the phenomenological component Disk with the physical relativistic reflection model $\texttt{relxill}$ (v2.6) 
\citep{Garcia2014, Dauser2014},
resulting in a total model defined as $\texttt{TBabs} * (\texttt{zpowerlw} + \texttt{relxill} + \sum_{j=1,2} \texttt{rdblur}_j * (\texttt{MYtorusL}_j + \texttt{MYtorusS}_j))$. 
In the \texttt{relxill} component, the photon index $\Gamma$ was tied to that of \texttt{zpowerlw} and allowed to vary. 
The inner disk radius ($R_{\rm in, Rel}$) was treated as a free parameter, assuming a fixed black hole spin parameter of $a = 0.99$. 
The emissivity indices were tied ($q_{1} = q_{2}$) and allowed to vary as a single relativistic emissivity index ($q_{\rm Rel}$). 
To account only for the reflected emission, the reflection fraction ($f_{\rm Ref}$) was fixed at $-1$. 
The inclination angle ($i_{\rm Rel}$) was tied to that of the MYTorus components and treated as a free parameter.
Other parameters, including ionization parameter ($\log \xi$), iron abundance ($A_{\rm Fe}$), and the model normalization ($N_{\rm Rel}$), were also allowed to vary during the fitting process.
This final physical model yielded a $C$-statistic of $4106.2$ for $4152$ d.o.f. and an AIC value of $4132.3$. 
The best-fit model, residuals, and parameters are presented in Fig.~\ref{Xray}(a), and Table~\ref{tab1}, with a detailed close-up of the Resolve spectrum shown in Fig.~\ref{Xray}(b).

Our spectral modeling reveals that the relativistically-blurred Fe-K$\alpha$ line and its associated reflection component are significantly required, in addition to the thermal Comptonization continuum and a narrow neutral Fe-K$\alpha$ line originating from distant matter such as the broad-line region and a dusty torus (Fig.~\ref{Xray}a). 
While previous studies using X-ray CCDs have discussed disk truncation in 3C120 \citep{Ballantyne2004, Kataoka2007,Lohfink2013}, those measurements were inherently limited by energy resolution; the detailed relativistic line profile was often degenerate with complex narrow Fe~XXV/XXVI lines \citep{Tombesi2014, Tombesi2017} and distant reflection.
Our observations with the X-ray microcalorimeter, Resolve, provide the first high-resolution spectroscopic confirmation of this relativistic line profile, yielding the most robust and precise constraint on the truncation radius to date. 
Notably, this high-resolution spectrum lacks clear ionized absorption lines in the Fe-K band (Fig.~\ref{Xray}b), potentially reflecting the transient nature of ultra-fast outflows.
Crucially, the profile of the relativistic Fe-K$\alpha$ line, governed by Doppler beaming and gravitational redshift, constrains the inner radius of the optically thick accretion disk to $R_{\rm in} = 17.4^{+6.3}_{-6.5}~R_{\rm g}$ ($1\sigma$ confidence level). 

\begin{figure*}[t]
\begin{center}
\includegraphics[width=16cm, angle=0]{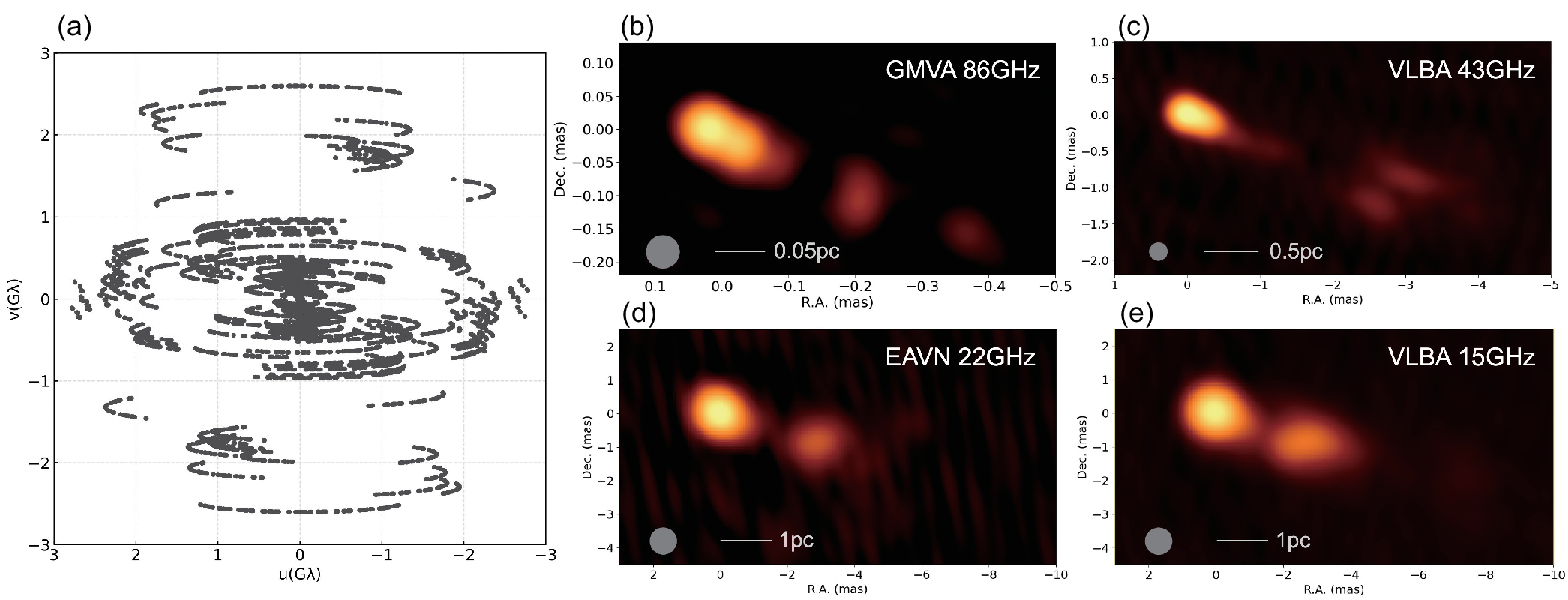}
\end{center}
\caption{(a)$(u,v)$ coverage of the GMVA observations in April 2025. (b) GMVA 86\,GHz image. (c) VLBA 43\,GHz image. (d) EAVN 22\,GHz image. (e) VLBA 15\,GHz image. The grey circle in the bottom-left corner of each panel (b)--(e) indicates the convolving beam size at each frequency.}\label{fig:vlbi}
\end{figure*}

\renewcommand{\arraystretch}{1.1}
\begin{table*}[htbp]
 \begin{center}
 \caption{Broken power-law model fit to jet collimation profile. Uncertainties are quoted at the $1\sigma$ level.}
 \label{tab:collimation}
  \vspace{0.5cm}
 \begin{tabular}{cccccc}

  \hline\hline
    $a_u$ & $a_d$ & $z_0$ & $R_{\rm jet, 0}$ & $\chi^2/{\rm d.o.f}$ \\ 
    & & ($R_{\rm g}$) & ($R_{\rm g}$) &  \\ \hline
   	$0.61\pm0.07$ & $1.13\pm 0.04$ & $(3.13\pm0.77)\times 10^5$ & $(7.23\pm1.72)\times 10^3$	& 1.18	\\	
  \hline\hline
  \end{tabular}
   \end{center}
  \end{table*}

\begin{figure*}[t]
\begin{center}
\includegraphics[width=11cm, angle=0]{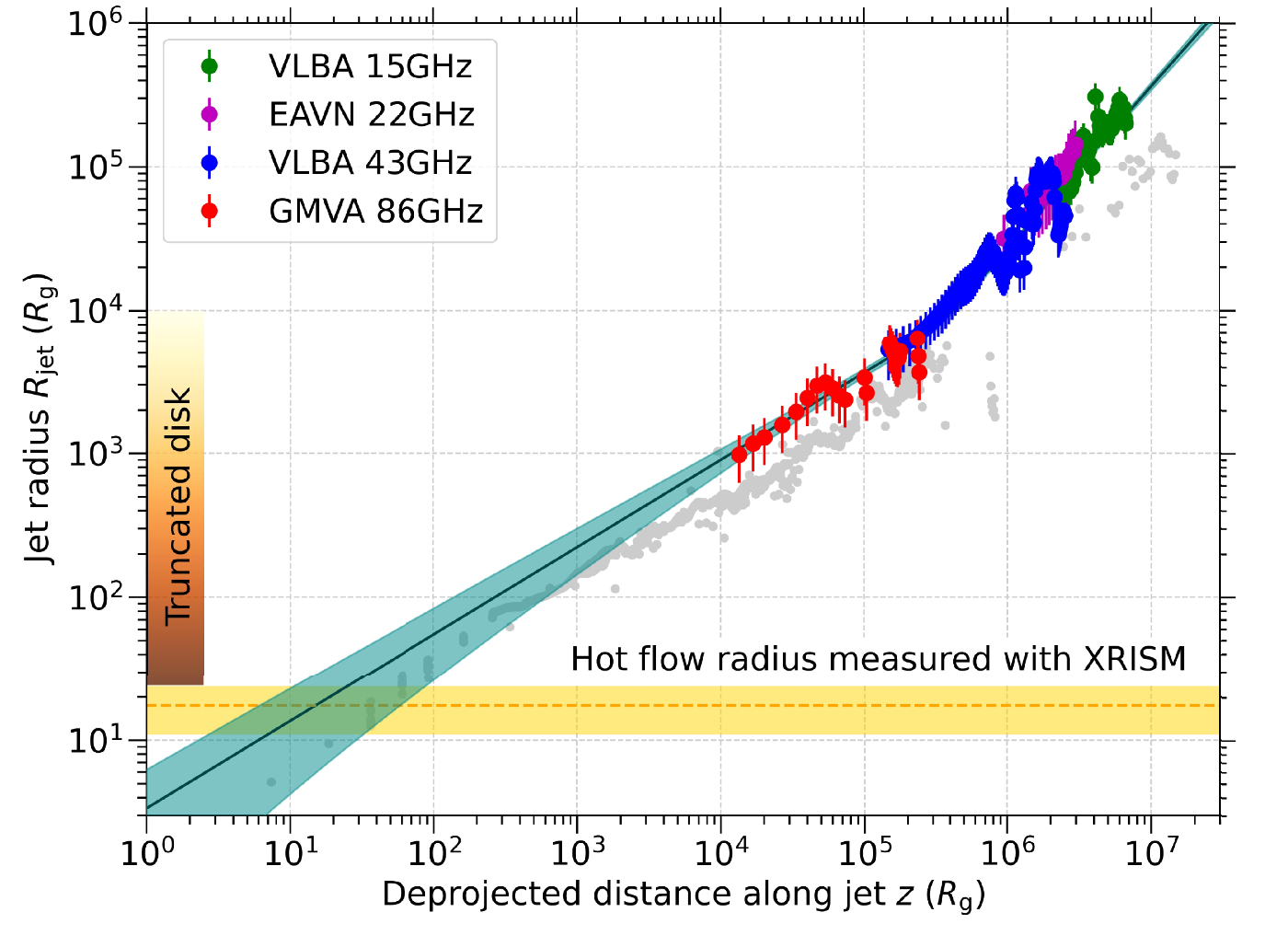}
\end{center}
\caption{\label{fig:collimation}
Jet radius of 3C120 as a function of deprojected distance from the radio core at 86\,GHz, both expressed in units of the gravitational radius. The red, blue, magenta and green data points correspond to measurements from the GMVA 86\,GHz, VLBA 43\,GHz, EAVN 22\,GHz and VLBA 15\,GHz images, respectively. The black solid line shows the best-fit broken power-law jet profile ($R_{\rm jet}(z) \propto z^{0.61\pm0.07}$ for $z<z_0$ and $R_{\rm jet}(z) \propto z^{1.13\pm0.04}$ for $z>z_0$ where $z_0 = (3.13\pm0.77) \times 10^5\,R_{\rm g}$) with the cyan-shaded region representing 1$\sigma$ uncertainty range of the fit. The horizontal gold-shaded band denotes the radius of the hot accretion flow ($R_{\rm in}=17.4^{+6.3}_{-6.5}~R_{\rm g}$) inferred from the XRISM spectral analysis (Fig.~\ref{Xray}). For reference, the jet profile of M87 \citep{Asada2012, Hada2024} is overlaid as grey data points. All quoted uncertainties represent the $1\sigma$ confidence level.}
\end{figure*}

To check how the results change with a different physical reflection model, we replaced \texttt{relxill} with \texttt{relxillCp} \citep{Garcia2018}, which assumes a thermal Comptonization component (\texttt{nthcomp}) as an incident continuum and allows the accretion disk density to vary.
Correspondingly, the primary continuum component was changed from \texttt{zpowerlw} to \texttt{nthcomp}, with the seed photon temperature fixed at 0.01~keV and a disk blackbody input spectrum. 
The resulting model is expressed as $\texttt{TBabs} * (\texttt{nthcomp} + \texttt{relxillCp} + \sum_{j=1,2} \texttt{rdblur}_j * (\texttt{MYtorusL}_j + \texttt{MYtorusS}_j))$. 
In this fit, the photon index $\Gamma$ was tied between \texttt{nthcomp} and \texttt{relxillCp} and allowed to vary, while their normalizations were varied independently.
The electron temperature $kT_{\rm e}$  was fixed at 75~keV, following \citet{Zdziarski2001}. 
The disk density $\log (N/\textrm{cm}^{-3})$ in \texttt{relxillCp} was allowed to vary, while all other parameters were set to the same conditions as in the \texttt{relxill} fit.
This model yields a comparable fit quality (C-statistic/d.o.f. = 4106.7/4151, AIC = 4134.8), with best-fit parameters consistent with those obtained using \texttt{relxill} (Table~\ref{tab1}).
In particular, the inner disk radius and inclination angle remain virtually unchanged: $R_{\rm in} = 17.4^{+5.9}_{-6.9}~R_{\rm g}$ and $i_{\rm Rel} = 31.3^{+0.9}_{-0.7}$ degrees ($1\sigma$ confidence level), with the disk density pegged at its lower hard limit $\log (N/\textrm{cm}^{-3}) = 15.0^{+3.4}_{-0.0P}$, where subscript $P$ indicates that the parameter was pegged at the model boundary. 
Furthermore, fixing $kT_{\rm e}$ at $30$~keV (e.g., \citealt{Rani2018})
produces virtually identical results (C-statistic/d.o.f. = 4107.7/4151, AIC = 4135.8; $R_{\rm in} = 17.3^{+5.8}_{-7.1}~R_{\rm g}$ and $i_{\rm Rel} = 31.3^{+0.9}_{-0.7}$ degrees). 
These results confirm that the derived inner disk radius and inclination angle are robust against the choice of the reflection model and the primary continuum electron temperature.

These results demonstrate that the thin disk does not reach the innermost stable circular orbit (ISCO), but is instead truncated at a significantly larger radius. 
This finding implies that the accretion flow interior to $\sim 20~R_{\rm g}$ undergoes a state transition into a hot, geometrically-thick flow, which is consistent with the relatively hard power-law photon index of $\sim 1.8$ expected from thermal Comptonization in such a hot flow, thereby providing the physical environment necessary for jet launching (Fig.~\ref{main}c).

In this modeling, the black hole spin parameter was fixed at $a=0.99$. Because the determined inner radius lies well outside the ISCO, the spectral profile lacks sensitivity to the spin, precluding its precise measurement. 
Indeed, we confirmed that refitting the spectra with a non-rotating black hole ($a=0$) yielded identical parameters and fitting quality. 

In Fig.~\ref{Xray}(b), we noted residual features at approximately 6.67~keV and 6.70~keV, which potentially correspond to the Fe~\textsc{xxv} emission lines reported in earlier studies \citep{Tombesi2017}. 
Because these features are weak, we confirmed that including a photoionized plasma model to account for them does not affect the best-fit parameters of the relativistically-blurred reflection component. 
A detailed analysis of these fine spectral features is beyond the scope of this letter and will be presented in a forthcoming publication.

\subsection{Jet Collimation Profile Analysis}

To explore how the inferred accretion flow geometry compares with that of the innermost jet regions, we further analyzed the GMVA 86~GHz, the VLBA 43/15~GHz and the EAVN 22~GHz observations data of 3C120 obtained contemporaneously with the XRISM session. 
Figure~\ref{main}(b) presents the GMVA 86~GHz image of 3C120 (rotated by 90 degrees on the sky), with the corresponding (u, v) coverage shown in Fig.~\ref{fig:vlbi}(a). 
Thanks to its ultra-high angular resolution of $\sim$50 microarcseconds ($\mu as$), the GMVA enables us to directly resolve the innermost jet regions down to $\sim$0.03\,pc scales in unprecedented detail, while the lower-frequency VLBA and EAVN images trace the extended jet emission further downstream (Fig.~\ref{fig:vlbi}b--e). The bright radio core at the apparent jet base is supposed to be the closest vicinity of the central engine where the bulk of X-ray emission is produced. The jet appears narrowly collimated and one-sided, indicative of strong Doppler boosting in the approaching jet. 

The jet collimation profile of 3C120 was derived from the VLBI images following procedures similar to those widely adopted in many previous studies (e.g., \citealt{Hada2018}). First, each image at GMVA 86\,GHz, VLBA 43\,GHz, EAVN 22\,GHz, and VLBA 15\,GHz was convolved with a circular Gaussian beam of 50\,$\mu$as, 250\,$\mu$as, 800\,$\mu$as and 800\,$\mu$as, respectively (Fig.~\ref{fig:vlbi}b--e). We then extracted transverse slices across the 3C120 jet every $\sim$one-fifth of the beam size along the jet axis (a position angle of 248 degrees), and each slice was fitted with a single Gaussian function. The jet width $W_{\rm jet}(z)$ at a distance $z$ was defined as a deconvolved size of the fitted Gaussian, $W_{\rm jet}(z) = (\Theta^2_{\rm Gauss} - \Theta^2_{\rm beam})^{1/2}$, where $\Theta_{\rm Gauss}$ and $\Theta_{\rm beam}$ denote the FWHM of the fitted Gaussian and convolving beam, respectively. We then defined the corresponding jet radius as $R_{\rm jet}(z) = W_{\rm jet}(z)/2$. As the measured collimation profile of 3C120 implies a gradual transition from a parabolic to a conical geometry, similar to that observed in other AGN jets, we model the $(z, R_{\rm jet})$ relation using the following broken power-law form: 
\medskip
\begin{equation}
R_{\rm jet}(z) = R_{\rm jet, 0}\,2^\frac{a_u-a_d}{2}\left(\frac{z}{z_0}\right)^{a_u}\left[1 + \left(\frac{z}{z_0}\right)^s\right]^{\frac{a_d-a_u}{s}},
\end{equation}
where $a_u$, $a_d$, $z_0$, $R_{\rm jet, 0}$ and $s$ are the upstream jet power-law index, downstream power-law index, (deprojected) transition distance, jet width at the transition distance and a parameter controlling the sharpness of the transition, respectively. Here we use a fixed value $s=10$ since the fit is insensitive to $s$, and the remaining four parameters are fitted to the data. The results of best-fit broken power-law model is summarized in Table~\ref{tab:collimation}, and Fig.~\ref{fig:collimation} presents the corresponding fit. Here we adopt a black hole mass of $M_{\rm BH} = 6.3\times 10^{7}\,M_{\odot}$ and a jet viewing angle 19 degrees. The jet distance in Fig.~\ref{fig:collimation} is measured from the 86\,GHz core.  
We find that the 3C120 jet is well described by a conical shape on large scales ($R_{\rm jet}(z)\propto z^{1.13\pm0.04}$; $1\sigma$ confidence level), while transitioning to a quasi-parabolic profile ($R_{\rm jet}(z)\propto z^{0.61\pm0.07}$; $1\sigma$ confidence level) within a deprojected distance of $\sim$$10^5$\,$R_{\rm g}$ from the jet origin. Such a parabolic-to-conical transition at $\sim$$10^{5-6}$\,$R_{\rm g}$ is consistent with that observed in other AGN jets, such as M87 \citep{Asada2012, Hada2024}, suggesting that the collimation and acceleration zone of the 3C120 jet is spatially resolved. By extrapolating the parabolic profile towards the innermost region, we find that the jet radius converges to $\lesssim 7~R_{\rm g}$ near the black hole horizon. 
This scale is comparable to, or even smaller than, the $\sim20~R_{\rm g}$ radius of the hot accretion flow identified by XRISM, and is consistent with a theoretical expectation for a jet base that is as narrow as, or even narrower than, the surrounding hot flow \citep{Liska2022}. 
We emphasize that this argument is further supported by two considerations. First, previous reports of superluminal motions up to $6.2\,c$ impose a strong constraint that the actual jet viewing angle is smaller than 19 degrees \citep{Casadio2015}, with detailed modeling studies favoring values of $\sim$10 degrees \citep{Hardee2005}.  Second, the radio core (the apparent jet origin in VLBI images) may be offset from the true jet origin due to synchrotron self-absorption (the core-shift effect; \citealt{Lobanov1998}). Although the VLBI data used in this study are not optimized for a reliable core-shift measurement (e.g., non-simultaneous multi-frequency data and without phase-referencing), our tentative core-shift estimate using the optically-thin jet components infers core-shifts of $\sim$0.04\,mas (86--43\,GHz), $\sim$0.06\,mas (43--22\,GHz), and $\sim$0.2\,mas (22--15\,GHz), respectively. Both effects would additionally shift the overall jet profile in Fig.~\ref{fig:collimation} towards larger downstream distances, thereby further reducing the extrapolated jet radius at the true origin. Therefore, these results provide compelling evidence that the relativistic jet in 3C120 does not originates from a cold, geometrically-thin accretion disk, but launched from a hot, geometrically-thick accretion flow in the immediate vicinity of the black hole, likely powered by the extraction of its rotational energy.

Notably, the overall jet width of 3C120 appears systematically wider than that of the M87 jet (Fig.~\ref{fig:collimation}). This trend is consistent with the trend suggested by \cite{Boccardi2021} and \cite{Yi2024}, where jets in highly accreting AGN tend to be systematically wider than those in weakly accreting AGN.

\section{Discussion: A Universal Engine Launching Relativistic Jets in Highly Accreting Black Holes}
\label{four}

Current theoretical frameworks, primarily informed by numerical simulations and observations of low-luminosity AGN such as M87, posit that relativistic jets are driven by the Blandford-Znajek mechanism, which requires
two essential components: a rapidly spinning black hole and a geometrically-thick accretion flow capable of creating and supporting large-scale strong magnetic fields \citep{Tchekhovskoy2011,Liska20}. 
In systems with high accretion rates, such as 3C120 ($L/L_{\rm Edd} \sim 0.1$--0.2), however, the environment is expected to be dominated by a geometrically-thin disk extending to the ISCO. 
Such thin disks are theoretically predicted to be inefficient at transporting magnetic flux towards the vicinity of the black hole \citep{Lubow1994, Guilet2012, Guilet2013} and collimating a relativistic jet \citep{Komissarov07,Lyubarsky09}. 
This created a paradox: According to standard models, high-accretion systems should suppress hot, geometrically-thick flow necessary to power and collimate a relativistic jet, even if the black hole possesses high spin.

Our observations provide a solution to this paradox by revealing a discrete geometric transition within the inner accretion flow. 
While previous studies using CCD-based instruments have suggested disk truncation in 3C120 \citep{Ballantyne2004, Kataoka2007, Lohfink2013}, those measurements often had limited energy resolution, making it difficult to unambiguously distinguish the relativistic Fe-K$\alpha$ line from complex absorption and/or distant reflection components. 
Here, the X-ray microcalorimeter onboard XRISM/Resolve provides the first high-resolution look at the Fe-K$\alpha$ profile, offering the most robust and precise measurement of the truncation radius to date. 
By clearly resolving the line shape, we demonstrate that the cold, geometrically-thin disk indeed truncates at $\sim 20~R_{\rm g}$, where a hot and geometrically-thick flow replaces it (Fig.~\ref{main}c). 
Substantial disk truncation at high accretion rates ($L/L_{\rm Edd} \sim 0.1$) has also been observed in microquasars (e.g., \citealt{Zdziarski2021}), suggesting that such a truncation is a common phenomenon of black hole accretion flows regardless of black hole mass.
Furthermore, the VLBI evidence showing the jet radius narrowing to a scale of $\lesssim 10~R_{\rm g}$ confirms that this hot flow successfully delivers magnetic flux to the immediate vicinity of the horizon (Fig.~\ref{main}c). 
The inner hot flow and the jet collimation therein provide the precise geometry required for the Blandford--Znajek mechanism. Our scenario is also consistent with the argument that the jet power of 3C120 is close to the Blandford--Znajek limit for a magnetically arrested disk (MAD; \citealt{Narayan2003}), which requires a pair-dominated jet with pairs created by radiation from the hot inner flow at the jet base \citep{Zdziarski2022}.

These results imply that the physics of jet launching is governed by a mechanism that cannot be fully captured by simple classifications of accretion regimes.
By undergoing a structural state transition at $\sim 20~R_{\rm g}$, 3C120 overcomes the inherent magnetic limitations of high accretion to satisfy the conditions for a relativistic jet powered by the central black hole and its collimation.
It is also important to test whether the inner hot flow is in a MAD state, which is capable of sustaining the strong magnetic fields required for the Blandford--Znajek process. The jet magnetic flux of 3C120 appears consistent with the maximum flux sustainable on the black hole horizon \citep{Zdziarski2022}, and the geometrically-thick hot flow found here can advect such a flux, although this is a necessary rather than sufficient condition. Establishing this state definitively remains challenging with the present results alone, and the dynamical properties of the flow will be key.
As shown in Appendix, the Xtend 0.5--10~keV light curve shows a possible quasi-periodic oscillation (QPO) with a frequency of $\sim 1.2\times 10^{-5}$~Hz, which could be explained by periodic magnetic eruption in a MAD predicted by \cite{Ripperda2022,sridhar25} or the Lense-Thirring precession of the hot inner flow (e.g., \citealt{Ingram2019}).
In the future, the X-ray timing properties will be important for further understanding of the dynamical state of the inner hot accretion flow and relativistic jet production. 

Our X-ray spectral modeling further reveals a geometric misalignment. 
The resulting inclination angle of the accretion disk, $i_{\rm rel} = 31.3^{+1.0}_{-0.4}$~degrees ($1\sigma$ confidence level; see Table~1) remains significantly larger than the value of $\sim 19$~degrees determined by the one-sided jet \citep{Casadio2015}. 
The observed discrepancy between the jet inclination ($\sim 19$~degrees) and the outer disk inclination ($\sim 31$~degrees) indicates a misalignment between the black hole spin axis and the angular momentum vector of the accretion disk. 
As GRMHD simulations have demonstrated (e.g., \citealt{Chatterjee25}), the jet direction is intrinsically tied to the black hole spin. 
The jet thus follows the black hole spin rather than the outer disk, as expected for a jet launched from the spin-aligned inner region.
Our observed geometric offset suggests that the inner accretion flow is tilted relative to the outer thin disk, a configuration that naturally induces Lense--Thirring precession of the inner hot flow around the black hole spin axis. 
The potential QPO identified in the X-ray flux variation (see Appendix) may be attributed to this precessional motion. 
While coherent precession may be suppressed in a prograde MAD due to electromagnetic alignment \citep{Fragile2023, Chatterjee25}, it could be sustained if the black hole possesses a retrograde spin~\citep{Gupta2026}. 
If the precession is completely suppressed, quasi-periodic magnetic flux eruptions with Comptonizing plasmoid formation \citep{Ripperda2022,sridhar25} offer an alternative scenario for the QPO.

The existence of truncated disks and inner hot flows has been suggested for some Seyfert galaxies, which are in the high-accretion regime without powerful radio jets (e.g., \citealt{Noda2016,Kubota2018}). 
If such a geometric transition is indeed a common feature across highly accreting systems, regardless of whether they launch jets or not, it implies that the presence of an inner hot flow is a necessary but not sufficient condition for powerful jet activity. 
While the main driver of the radio-loud/radio-quiet dichotomy and the dominant role of black hole spin have been actively debated since \citet{Sikora2007}, our findings provide key insight into this long-standing issue.
Because 3C120 possesses the geometrically-thick environment required to support large-scale magnetic fields, our results suggest that black hole spin serves as the decisive factor \citep{Tchekhovskoy10}.
This indicates that once disk truncation creates a magnetically favorable environment, the ability of the system to tap into the rotational energy of the black hole becomes the primary engine of the relativistic jet, providing a unified spin-driven framework for the disk--jet connection across the AGN population.

\section{Conclusions}

In this work, we presented the first high-resolution X-ray spectroscopic observation of the relativistically-broadened Fe-K$\alpha$ line and reflection component in the broad-line radio galaxy 3C120 using XRISM, establishing that the cold, geometrically-thin accretion disk truncates at $\sim 20~R_{\rm g}$ and transitions into a hot, geometrically-thick inner flow. 
Supported by independent VLBI evidence of jet collimation within $\lesssim 10~R_{\rm g}$ on the horizon scale, this discovery offers a direct geometric solution to the longstanding theoretical paradox regarding how relativistic jets are launched and collimated in highly accreting systems where a standard accretion disk dominates. 
This study demonstrates that disk truncation provides the necessary physical condition to transport magnetic flux and trigger the Blandford--Znajek process. 
Furthermore, a marginal QPO candidate in the XRISM/Xtend light curve (discussed in Appendix) offers a supplementary temporal perspective consistent with dynamic magnetic flux eruptions in a MAD or coherent precession, hinting at a highly magnetized inner engine. 
These results highlight the pivotal role of coordinated high-resolution X-ray spectroscopy and VLBI imaging in unraveling black hole magnetohydrodynamics, and demonstrate how inner structural transitions establish magnetically favorable sites for spin-powered jets.

\section*{Acknowledgments}
We thank the anonymous referee for constructive comments and suggestions that improved the paper.
This work was supported by JSPS KAKENHI grant numbers 
23K20239, 24K00672, 25H00660 (H.N.), 26K21725 (K.H.). K.H. is supported by Daiko Foundation (grant J0SE807004). This work was supported in part by a University Research Support Grant from the National Astronomical Observatory of Japan (NAOJ), and by Grant-in-Aid for Outstanding Research Group Support Program in Nagoya City University Grant Number 2530002. I.C. is supported by the Glocal University 30 Project Fund of Kyungpook National University in 2026, and Global - Learning \& Academic research institution for Master's, PhD students, and Postdocs (LAMP) Program of the National Research Foundation of Korea (NRF) grant funded by the Ministry of Education (No. RS-2023-00301914). 
This research has made use of data obtained with the Global Millimeter VLBI Array (GMVA), which consists of telescopes operated by the \"{u}r Radioastronomie (MPIfR), IRAM, Onsala Space Observatory, Mets\"{a}hovi Radio Observatory, Yebes Observatory, the Green Bank Observatory, and the Very Long Baseline Array (VLBA). The Greenland Telescope (GLT) is operated by the Academia Sinica Institute of Astronomy and Astrophysics (ASIAA) and the Smithsonian Astrophysical Observatory (SAO). The GMVA data were correlated at the VLBI correlator of MPIfR in Bonn, Germany. This study makes use of VLBA data from the VLBA-BU Blazar Monitoring Program (BEAM-ME and VLBA-BU-BLAZAR; 
http://www.bu.edu/blazars/BEAM-ME.html, funded by NASA through the Fermi Guest Investigator Program. This research has made use of data from the MOJAVE database that is maintained by the MOJAVE team~\citep{Lister2018}. The VLBA is an instrument of the National Radio Astronomy Observatory. The National Radio Astronomy Observatory is a facility of the National Science Foundation operated by Associated Universities, Inc. This work made use of the East Asian VLBI Network (EAVN), which is operated under cooperative agreement by the National Astronomical Observatory of Japan (NAOJ), Korea Astronomy and Space Science Institute (KASI), Shanghai Astronomical Observatory (SHAO), Xinjiang Astronomical Observatory (XAO), Yunnan Observatories (YNAO), National Astronomical Research Institute of Thailand (Public Organization) (NARIT), and National Geographic Information Institute (NGII), with the operational support by Ibaraki University (for the operation of Hitachi 32 m and Takahagi 32 m), Yamaguchi University (for the operation of Yamaguchi 32 m) and Kagoshima University (for the operation of VERA Iriki antenna). The Nanshan 26 m radio telescope (NSRT) is operated by the Urumqi Nanshan Astronomy and Deep Space Exploration Observation and Research Station of Xinjiang. 

\appendix

\section{X-ray Timing and PSD Analysis}

The Xtend light curves were extracted in the 0.5--10~keV energy range with 5 ksec time binning (Fig~\ref{qpo}a). 
The background count rate, estimated from a source-free region of identical area, was found to be less than $\sim 0.2$~counts/s. 
This is negligible compared to the source intensity, ensuring that the detected variability is intrinsic. 
The light curve shows a fractional variability amplitude $\sim10\%$ on timescales of $\sim 100$~ksec. 
While the light curve exhibits a quasi-periodic modulation, we note that only 4--5 cycles were captured, necessitating a cautious interpretation.

Based on a previously reported QPO frequency at $\sim 7.1 \times 10^{-6}$~Hz \citep{Agarwal2021}, we defined a search window from $f_{\rm min} = 6 \times 10^{-6}$ Hz to $f_{\rm max} = 2 \times 10^{-5}$~Hz to allow for possible frequency drift and finite QPO width.  
The lower bound was chosen to retain sufficient low-frequency bins for constraining the underlying red-noise continuum.  
The Power Spectral Density (PSD) was calculated from the light curve using the Stingray software package (https://docs.stingray.science/en/stable/)
, with fractional rms normalization. 
The observed PSD was modeled as a combination of a red-noise component, described by a power-law $P(f) = N f^{-\alpha}$, and a constant Poisson noise floor $C_{\rm fit}$. 
The parameters $N$ and $\alpha$ were obtained by fitting the observed PSD over the frequency range of $3 \times 10^{-6}$ to $10^{-4}$ Hz. 
To ensure the robustness of the continuum estimation, we excluded the lowest-frequency bin from the analysis, as it is subject to large statistical uncertainties and red-noise leakage effects. 
Additionally, we masked the pre-defined QPO search window ($6 \times 10^{-6}$ to $2 \times 10^{-5}$~Hz) during the fitting process to prevent the potential signal from biasing the red-noise continuum estimation. 
The constant Poisson noise floor $C_{\rm fit}$ was independently estimated from the median power in the higher frequency regime ($> 4 \times 10^{-5}$ Hz).

\begin{figure*}[t]
\setcounter{figure}{0}
\renewcommand{\thefigure}{A\arabic{figure}}
\begin{center}
\includegraphics[width=15cm, angle=0]{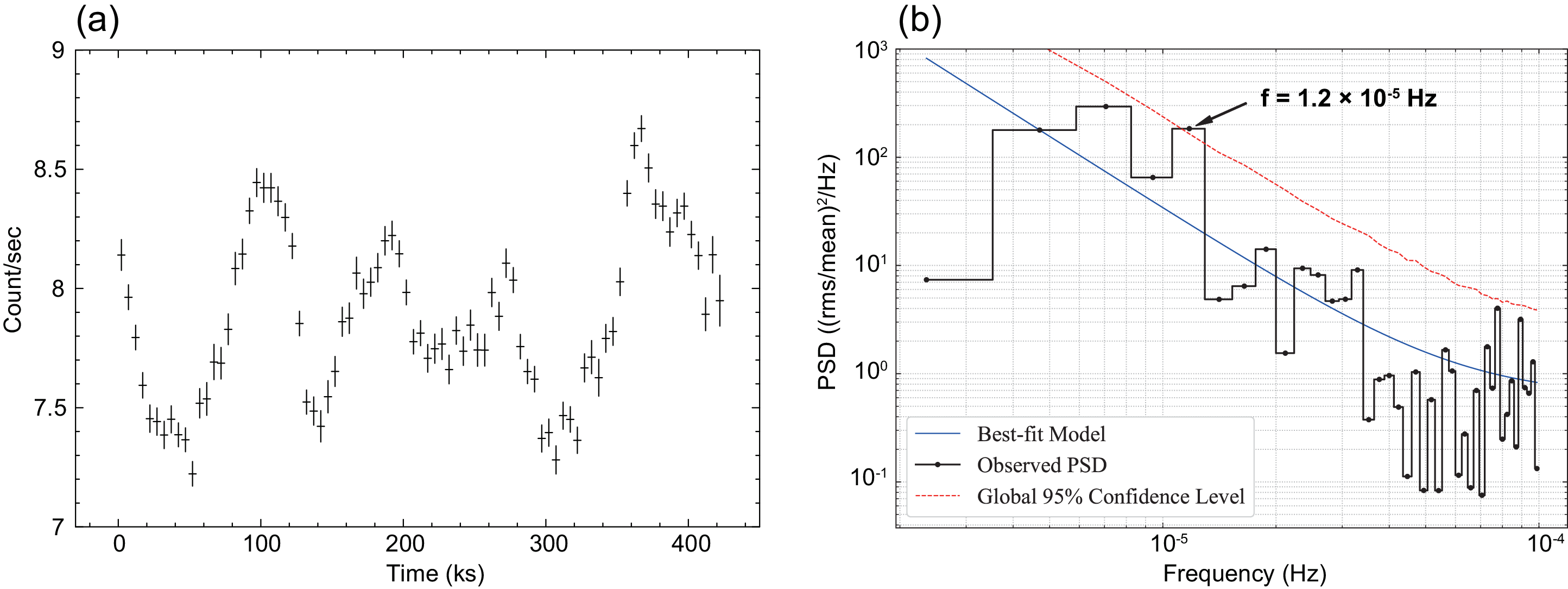}
\end{center}
\caption{(a) XRISM Xtend light curve in the 0.5--10~keV band, showing the temporal evolution of the X-ray flux. (b) PSD derived from the light curve, exhibiting a significant peak at approximately $1.2 \times 10^{-5}$~Hz. The solid blue line represents the best-fit model (red noise plus Poisson noise), and the dashed red line indicates the global 95\% confidence level. The detection of the peak above this threshold confirms its statistical significance against the underlying red-noise continuum.}
\label{qpo}
\end{figure*}

To assess the significance of power excesses against the red-noise background, we performed 10,000 Monte Carlo simulations following the statistical framework for red-noise testing established by \cite{Vaughan2005}. 
To account for red-noise leakage and ensure frequency-dependent consistency, each simulation was generated with a duration ten times longer than the observation and then truncated to the original length. 
The simulated PSDs were normalized to ensure the ensemble average matched the best-fit theoretical model. 
We accounted for the ``Look-Elsewhere Effect'' (LEE) within our primary frequency range of interest ($6 \times 10^{-6}$ to $2 \times 10^{-5}$ Hz) by calculating a global 95\% confidence threshold. 
The global significance was determined by adjusting the local $p$-value such that the overall probability of a false-positive detection across all searched bins remains below 0.05.

The resulting PSD is presented in Fig~\ref{qpo}(b). 
Our analysis revealed a power excess centered at a frequency of $\sim 1.2 \times 10^{-5}$~Hz, corresponding to a periodicity of approximately $1$~days. 
The frequency of this feature is slightly higher than that of the periodicity previously reported in a NuSTAR study \citep{Agarwal2021}. 
While we assumed a simple power-law for the underlying continuum in our simulations, we acknowledge that the significance of this feature could be sensitive to the choice of the noise model. 
Given that only a limited number of cycles were covered during the observation, we report this detection as a marginal result. 
Further investigations with longer baseline observations will be required to definitively distinguish this feature from complex red-noise processes.

\section{Interpretation of possible QPO}

The XRISM/Xtend light curve reveals a marginal periodic modulation at approximately $1.2 \times 10^{-5}$~Hz, providing an independent temporal constraint on the inner accretion flow. 
One natural interpretation for such low-frequency modulations is the Lense--Thirring precession of the hot inner flow, as widely discussed in X-ray binaries~\citep{Ingram2019}. 
The Lense--Thirring precession frequency for a test particle is given by
\begin{equation}
\nu_{\rm LT} = \frac{c}{2\pi R_{\rm g}}\frac{2a}{r^3},
\end{equation}
where $r = R/R_{\rm g}$ is the dimensionless radius and $a$ is the dimensionless black hole spin parameter. 
Substituting the spectrally constrained inner radius of $r = 17.4^{+6.3}_{-6.5}$, $a = 0.99$, and the black hole mass of 3C120 ($M_{\rm BH} \sim 6.3 \times 10^7~M_\odot$, which determines $R_{\rm g} = G M_{\rm BH}/c^2$) yields an expected frequency range of $\nu_{\rm LT} \sim 8 \times 10^{-8} \text{ -- } 8 \times 10^{-7}~{\rm Hz}$.
This predicted range is more than an order of magnitude lower than the observed modulation. 
Although a smaller precession radius concentrated well inside the disk truncation radius could elevate the expected frequency to match the observed value, recent numerical simulations suggest that such coherent hydrodynamic precession is difficult to maintain if the inner flow has entered a MAD phase. 
This is because the strong magnetic fields associated with the jet exert an electromagnetic torque that dynamically aligns the flow with the black hole spin axis \citep{Fragile2023,Chatterjee25}.
However, recent GRMHD simulations demonstrate that if the black hole possesses a retrograde spin, the altered electromagnetic torque allows coherent Lense--Thirring precession of the MAD to be sustained \citep{Gupta2026}.

Alternatively, this periodic signal may originate from intrinsic magnetohydrodynamic processes within the MAD itself, rather than simple rigid body precession. 
Recent high-resolution simulations demonstrate that MADs undergo periodic magnetic reconnection events near the event horizon, leading to the formation of plasmoids and associated flux eruptions \citep{Ripperda2022,sridhar25}. 
These events manifest as quasi-periodic oscillations in the electromagnetic emission and are intrinsically linked to the regulation of jet power. 
Given that our results already place 3C120 in a magnetically favorable state through disk truncation, the observed $1.2 \times 10^{-5}$~Hz signal may represent the first direct detection of such magnetic flux dynamics in a high accretion system. 
This would imply that the oscillation is a signature of the engine itself, periodically releasing Comptonizing plasmoids associated with a highly-magnetized jet, reconciling the observed timing signal with the magnetohydrodynamic requirements of a jet launching site.

\bibliography{3c120}

\end{document}